\documentstyle[floats,prb,aps,epsf]{revtex}
\begin{document}
\draft
\title{Interface resistance in ferromagnet/superconductor junctions}
\author{A. A. Golubov}
\address{Department of Applied Physics, University of Twente, \\
P.O.Box 217, 7500 AE Enschede, The Netherlands}
\maketitle

\begin{abstract}
Results of theoretical study of spin-polarized tunneling in
ferromagnet/superconductor junctions are presented. Spin and charge currents
are calculated as a function of applied voltage and spin polarization in a
ferromagnet. The model takes into account the splitting of different spin
subbands in a ferromagnet and impurity scattering in the contact. The excess
resistance of an FS contact due to the charge-imbalance in a superconductor
is calculated for the first time. The results have implications for
spin-coupled magnetoresistance in ferromagnet/superconductor contacts and
for measuring spin polarization in ferromagnets.
\end{abstract}

\section{Introduction}

The rapidly emerging field of spin polarized transport is based on the
ability of a ferromagnetic metal to conduct and accumulate spin-polarized
currents \cite{Tedrow,Prinz,Johnson}. Spin polarized transport between
ferromagnets (F) and superconductors (S) received considerable attention
recently because of new physical phenomena and potential device
applications. An introduction of the hybrid structures based on a
combination of ferromagnetic and superconducting materials are not only
interesting from a fundamental point of view but can bring further
advantages for devices \cite{Prinz,Johnson}. In particular, spin
accumulation effects in superconductors may play an important role because
of a number of reasons. First, due to the gap in the excitation spectrum,
the spin diffusion length in a superconductor can become quite long at low
temperatures \cite{Tedrow,Hersh}. Second, spin accumulation can take place
at a FS interface since spin-polarized current in a ferromagnet has to be
transformed into spinless supercurrent in a superconductor \cite
{Falko,Jedema}. An important step in the quantitative analysis of spin
accumulation and spin injection in superconductors is the knowledge of the
dependence of the resistance of a FS interface on the spin polarization in a
ferromagnet.

Furthermore, it was recently demonstrated that a combination of F and S
metals can be used advantageously for measuring spin polarization in
metallic ferromagnets either by measuring $T_c$ of FS multilayers \cite
{Aarts}, or by directly measuring the resistance of FS point contacts \cite
{Soul,Bur}. So far, theoretical studies of the FS contact resistance were
limited by calculations for ballistic FS contacts \cite
{Soul,Bur,JB,Kash,Maz,Merill,Zutic}. It was argued in \cite{Maz} that the
effects of impurity scattering are quite important in spin-polarized
tunneling since the degree of spin polarization is defined differently in
ballistic and diffusive contacts. However, no quantitative calculations on
the effects of disorder have been published up to now. Moreover, the
contribution of the contact resistance is most easily measured only in a
point contact geometry, while for larger area contacts the contribution of
an interface becomes rather small. The purpose of the present paper is
twofold. First, we extend the theory in order to include the effects of
impurity scattering in a contact. Second, we argue that an additional
sensitive probe for spin polarization is the excess resistance $R_{ex}$ of a
FS contact. This resistance is due to penetration of an electric field into
a superconductor over macroscopically large charge-imbalance relaxation
length $\lambda _Q$ and may exceed the direct interface resistance. We show
that the magnitude of $R_{ex}$ is sensitive to spin polarization in a
ferromagnet and provide an estimate for this effect.

\section{Ballistic FS contact}

We start from the derivation of the general expression for the conductance
of a FS contact in the absence of impurity scattering (ballistic case). We
consider the atomically sharp interface barrier at $x=0$ separating F metal (%
$x<0$) and S metal ($x>0$), modeled by a potential $U(r)=H\delta (x)$ and
arbitrary relation between Fermi velocities in F and S, $v_{F\uparrow },$ $%
v_{F\downarrow }$ and $v_{Fs}$. Here $H$ is the barrier strength parameter, $%
v_{F\uparrow ,\downarrow }=\sqrt{2E_{F\uparrow ,\downarrow }/m}\equiv \sqrt{%
2E_F(1\pm h)/m}$, $v_{Fs}=\sqrt{2E_{Fs}/m}$, where $E_{ex},E_F$ and $E_{Fs}$
are respectively exchange energy in a ferromagnet and Fermi energies in F
and S metals, the indices $\uparrow ,\downarrow $ refer to the spin subbands
and $h=E_{ex}/E_F$ denotes the dimensionless spin polarization in a
ferromagnet. We assume that the effective electronic masses $m_F$, $m_S$ are
equal to the free electron mass $m_e$, the mean free path is larger than the
size of the contact and the pair potential is approximated by the step
function $\Delta (x)=\Delta (T)\theta (x)$, $\Delta (T)$ being the bulk pair
potential in a superconductor.

Charge and spin currents can be calculated within the framework of the BTK
approach \cite{BTK}, i.e. considering explicitly Andreev and normal
reflections at the FS interface and taking into account that an incoming
electron and an Andreev reflected hole occupy opposite spin subbands \cite
{JB}. The electron- and hole--like excitations are represented by
two-component wave functions, which obey the Bogoliubov de-Gennes equations.
An electron, incoming from the ferromagnet F into the superconductor S, is
described by a plane wave with a wave vector $k_{\uparrow }^{+}$%
\begin{equation}
\psi _{inc}=\left( 
\begin{array}{c}
1 \\ 
0
\end{array}
\right) e^{ik_{x\uparrow }^{+}x}.  \label{in}
\end{equation}
Due to the four-fold degeneracy of an excitation in a superconducting state,
the electron is partially reflected into F as an electron with the opposite
wave vector $-k_{x\uparrow }^{+}$ or as a hole $k_{x\downarrow }^{-}$

\begin{equation}
\psi _{refl}=a\left( 
\begin{array}{c}
0 \\ 
1
\end{array}
\right) e^{ik_{x\downarrow }^{-}x}+b\left( 
\begin{array}{c}
1 \\ 
0
\end{array}
\right) e^{-ik_{x\uparrow }^{+}x}  \label{ref}
\end{equation}
and partially transmitted into S without branch crossing $k_{sx}^{+}$, or
with branch crossing $k_{sx}^{-}$%
\begin{equation}
\psi _{trans}=c\left( 
\begin{array}{c}
u \\ 
v
\end{array}
\right) e^{ik_s^{+}x}+d\left( 
\begin{array}{c}
v \\ 
u
\end{array}
\right) e^{-ik_s^{-}x}.  \label{tr}
\end{equation}
Here $k_{x\uparrow }^{\pm },$ $k_{x\downarrow }^{\pm }$ and $k_s^{\pm }$ are
the projections of the Fermi wave vectors in two spin sibbands and in a
superconductor on the direction $x$ normal to the contact plane, index $+$ ($%
-$) refers to electron- or hole-like quasiparticles.

The amplitudes $a,b,c$ and $d$ have to be determined from the matching
conditions for $\Psi _F=\psi _{inc}+\psi _{refl}$ and $\Psi _S=\psi _{trans}$
at the interface, $x=0$

\begin{eqnarray}
\Psi _F(0) &=&\Psi _S(0),  \label{bc} \\
\frac{d\Psi _F(0)}{dx}-\frac{d\Psi _S(0)}{dx} &=&2m_eH\Psi _F(0)/\hbar . 
\nonumber
\end{eqnarray}
Using these conditions we find the Andreev and normal reflection
coefficients, $A=\left| a\right| ^2$and $B=\left| b\right| ^2$, and the
transmission coefficients with or without branch crossing, $C=\left|
c\right| ^2$ and $D=\left| d\right| ^2$, respectively, which determine
charge and spin currents in a ballistic FS contact.

Charge current in a FS contact is given by 
\begin{equation}
I=\frac e{2\pi \hbar }\sum_{\uparrow ,\downarrow }P_{\uparrow ,\downarrow
}\int \frac{d^2k_{\parallel }}{(2\pi )^2}\int_{-\infty }^{+\infty }\left[ 1+%
\frac{k_{F\downarrow ,\uparrow }}{k_{F\uparrow ,\downarrow }}A_{\uparrow
,\downarrow }(E)-B_{\uparrow ,\downarrow }(E)\right] \left[
f(E+eV)-f(E)\right] dE,  \label{cur}
\end{equation}
where $P_{\uparrow \downarrow }=(E_F\pm E_{ex})/2E_F=(1\pm h)/2$, $%
v_{F_x\uparrow ,\downarrow }$ is the projection of the Fermi velocity on the
direction $x$, $f(E)$ is the Fermi distribution function, and $k_{\parallel }
$ is the component of the Fermi momentum parallel to the junction plane,
which is conserved for each individual scattering process. The ratio $%
k_{F\downarrow ,\uparrow }/k_{F\uparrow ,\downarrow }$ provides the
normalization of the total probability current, taking into account that
Andreev scattering involves different spin subbands.

In the limit of low temperatures $T\ll T_c$, we arrive the following
expression for the charge conductance of a ballistic FS contact at the
subgap bias voltage $eV<\Delta (T)$

\begin{equation}
G_{FS}=2G_0T_{\uparrow }T_{\downarrow }\frac{(1+\alpha ^2)P_{\downarrow
}(v_{F\uparrow }+v_{F\downarrow })/v_{Fs}}{(1-r_{\uparrow }r_{\downarrow
})^2+\alpha ^2(1+r_{\uparrow }r_{\downarrow })^2}.  \label{Gfs}
\end{equation}
Here $G_0=e^2k_F^2S/4\pi ^2\hbar $ is the normal state (Sharvin) conductance
of the contact, $S$ is the contact area, $T_{\uparrow }$ and $T_{\downarrow }
$ are the transmission probabilities for scattering from the spin up(down)
subband into a superconductor

\begin{equation}
T_{\uparrow }=\frac{4v_{F\uparrow }v_{Fs}}{4Z^2+(v_{F\uparrow }+v_{Fs})^2},%
\text{ }T_{\downarrow }=\frac{4v_{F\downarrow }v_{Fs}}{4Z^2+(v_{F\downarrow
}+v_{Fs})^2},\text{ }  \label{T12}
\end{equation}

\begin{equation}
r_{\uparrow ,\downarrow }=\sqrt{1-T_{\uparrow ,\downarrow }^{}},\text{ }%
Z=H/\hbar v_{Fs},\text{ }\alpha =\sqrt{\Delta ^2(T)-(eV)^2}/eV.  \label{Z}
\end{equation}

In relevant limits the expressions (\ref{Gfs})-(\ref{Z}) agree with the
results derived in \cite{Soul,Bur,JB,Kash,Merill,Zutic}, while the
advantage of the representation (\ref{Gfs})-(\ref{Z}) is, that the charge
conductance is directly expressed in terms of the individual probabilities $%
T_{\uparrow ,\downarrow }^{}$ and therefore is particularly suitable for
consideration of the impurity scattering, as explained in the next section.
It follows from eq.(\ref{Gfs}) that in a NS contact ($T_{\uparrow
}=T_{\downarrow }=T$ ) at zero bias the BTK result $G_{NS}=2G_0T^2/(2-T)^2$
is recovered, which yields the {\it conductance doubling} $G_{NS}=2G_0$ for
a fully transmissive contact, $T=1$. This conductance enhancement is
suppressed in a FS contact when $T_{\uparrow }\neq T_{\downarrow }$ due to
spin polarization in a ferromagnet.

It is straightforward to extend the expressions (\ref{Gfs})-(\ref{Z}) to the
regime of high bias voltage, $eV>\Delta (T)$. The results of numerical
calculations of the dependence of charge current and zero-bias conductance
are presented in Figs.1-3 for various values of spin polarization. It is
seen that the zero-bias charge conductance is quite sensitive to the spin
polarization $h=E_{ex}/E_F$. Fig.3 shows that for small values of $h$ this
dependence is linear, which reflects the simple fact that the number of
transmitted electronic modes scales like $1-h$ due to spin reversal by the
Andreev reflection. At $h$ close to 1 a nonlinearity appears in the $G_{FS}$
vs $h$ dependence due to the dependence of the transmission coefficients $%
T_{\uparrow ,\downarrow }^{}$ on $h$ (\ref{T12}).

Spin current in a FS contact is given by 
\begin{equation}
I=\frac e{2\pi \hbar }\sum_{\uparrow ,\downarrow }P_{\uparrow ,\downarrow
}\int \frac{d^2k_{\parallel }}{(2\pi )^2}\int_{-\infty }^{+\infty }\left[ 1-%
\frac{k_{F\downarrow ,\uparrow }}{k_{F\uparrow ,\downarrow }}A_{\uparrow
,\downarrow }(E)-B_{\uparrow ,\downarrow }(E)\right] \left[
f(E+eV)-f(E)\right] dE  \label{spin}
\end{equation}
Difference in the sign of the contributions of the Andreev coefficients $%
A_{\uparrow \downarrow }(E)$ in eqs. (\ref{cur}) and (\ref{spin}) reflects
the fact that, while an Andreev reflected hole carries charge in the same
direction as an incoming electron, it carries spin in the opposite direction.

The low temperature spin current vanishes at subgap voltages, while at $%
eV>\Delta (T)$ the spin conductance $G_{FS}^{(s)}$ is given by the following
expression

\begin{eqnarray}
G_{FS}^{(s)} &=&\frac{4G_0\beta }{\left[ 1+\beta -r_{\uparrow }r_{\downarrow
}(1-\beta )\right] ^2}\left[ \frac{T_{\uparrow }v_{F\uparrow }}{2v_{Fs}}+%
\frac{T_{\downarrow }P_{\uparrow }v_{F\downarrow }}{2v_{Fs}}\right] ,
\label{spin1} \\
\beta &=&\sqrt{(eV)^2-\Delta ^2(T)}/eV.
\end{eqnarray}

From eq.(\ref{spin1}) one can calculate numerically the spin conductance as
a function of bias voltage and spin polarization. Figs.1, 2 show the results of
numerical calculations of the dependences of the low temperature spin
conductance on the polarization in a ferromagnet for two different values of
the barrier strength parameter, $Z=0$ and $Z=1$.

\section{Diffusive FS contact}

In the previous section the case of a ballistic FS contact was considered,
when the contact size is smaller than the electronic mean free path. However
the latter condition is not always fulfilled in experiments, and it is
therefore of interest to evaluate the effect of impurity scattering in a
contact. As a model for a FS contact we consider two bulk reservoirs (S and
F), which in addition to the interface potential $U(r)=H\delta (x)$ are
separated by the scattering region (a diffusive conductor) with a size
smaller than the electronic mean free path.

The expressions derived above are particularly suitable for the introduction
of impurity scattering, since they allow straightforward application of the
scattering theory. According to this theory (see \cite{Dor,Naz,Been} and
references therein) any diffusive conductor having size larger than the
electronic mean free path is characterized by universal distribution of
transmission eigenvalues $t$ over different channels. An average conductance
of a diffusive metal is then given as a sum of contributions of those
channels, each having the conductance $G_0=e^2/2\pi \hbar $

\begin{equation}
G_\sigma =\frac{G_{N\sigma }}{G_0}\int_0^1g_\sigma (t)\rho (t)dt,
\label{dif}
\end{equation}
where $\sigma =\uparrow ,\downarrow $ is the spin index, $g_\sigma $ is the
conductance of a channel with transmission coefficient $t$, $G_{N\sigma
}=e^2N_\sigma (0)D_\sigma $ is the normal state conductance per spin
direction, $N_\sigma $ is the density of states at the Fermi level. The
expression (\ref{dif}) is valid when the impurity scattering does not mix
different spin directions.

Function $\rho (t)$ is the universal distribution function of transmission
eigenvalues for different channels given by \cite{Dor} 
\begin{equation}
\rho (t)=\frac 1{2t\sqrt{1-t}}  \label{distr}
\end{equation}
and does not depend on microscopic parameters of a diffusive conductor. Eq.(%
\ref{distr}) shows that the transmission eigenvalues have a bimodal
distribution with a peak at unit transmission and a peak at exponentially
small transmission.

As a model for the diffusive SF contact we consider two scattering regions
in series: an incoming electron is first transmitted through the diffusive
region with probability $t$, then crosses the FS interface with probability $%
T_{\uparrow }$. In turn, an Andreev-reflected hole is first scattered by the
interface (probability $T_{\downarrow }$), then by the diffusive region. The
probabilities of these two-step processes $T_{1,2}$ are given by the
expression

\begin{equation}
T_{1,2}=\frac{tT_{\uparrow ,\downarrow }}{t+T_{\uparrow ,\downarrow
}-tT_{\uparrow ,\downarrow }},  \label{Teff}
\end{equation}
which follows from averaging over transmission resonances between two
scattering regions, assuming that all relevant distances exceed the
electronic wave-length.

The charge conductance in a diffusive FS contact is given by the expression
eq.(\ref{Gfs}) in which the probabilities $T_{\uparrow ,\downarrow }$ should
be substituted by the probabilities $T_{1,2}$ of the two-step scattering
processes described above. Here we present the result for low temperatures
and $eV<\Delta (T)$

\begin{equation}
G_{FS}=G_N\int_0^1\frac{T_1T_2(1+\alpha ^2)P_{\downarrow }(v_{F\uparrow
}^2/v_{Fs}^2+1)}{(1-r_1r_2)^2+\alpha ^2(1+r_1r_2)^2}\frac{dt}{t\sqrt{1-t}},
\label{Gfsd}
\end{equation}
where the probabilities $T_{1,2}$ are given by eq.(\ref{Teff}), $r_{1,2}=%
\sqrt{1-T_{1,2}}$, $\alpha =\sqrt{\Delta ^2(T)-(eV)^2}/eV$ and $%
G_N=2e^2N(E_F)D(E_F)$ is the conductance of a contact in the unpolarized
state.

In the NS case with a transparent interface (Z=0) and $v_{F\uparrow
}=v_{F\downarrow }=v_{Fs}$ the above expression at $V=0$ yields 
\begin{equation}
G_{NS}(V=0)=G_N\int_0^1\frac{tdt}{(2-t)^2\sqrt{1-t}}\equiv G_N,  \label{AVZ}
\end{equation}
i.e. we reproduce the well known result that the zero-bias conductance of
the diffusive contact $G_{NS}=$ $G_N$, in contrast to the ballistic case
when $G_{NS}=2G_N$, first obtained by Artemenko,Volkov and Zaitsev \cite{AVZ}
by a different method.

Fig.4 shows the results of numerical calculations of the dependence of the
zero-bias conductance of a disordered FS contact vs spin polarization.
It is seen by comparison of Figs.3 and 4, that assuming ballistic transport in a 
FS contact one can overestimate the spin polarization in a ferromagnet. The
results presented here correspond however to the strong scattering regime.
For a more quantitative comparison with experiments the model should be
further extended to the regime of arbitrary scattering strength.

\section{ Excess resistance}

So far we have taken into account both the interface and impurity scattering
in the contact, but neglected the contribution of an electric field
penetrating a superconductor. The latter can be indeed neglected in a point
contact geometry, while it becomes important in planar contacts, in
particular close to $T_c$, when an electric field penetrates into a
superconductor over the macroscopically large charge-imbalance relaxation
length $\lambda _Q$ \cite{Hsiang}.

The corresponding contribution to the boundary resistance of a FS contact
can be calculated by the generalization of the approach of \cite{BTK,Hsiang}
valid for a clean superconductor. Excess resistance $R_{ex}$ is given 
\begin{equation}
R_{ex}=F\lambda _Q\rho _s/S.  \label{Rex}
\end{equation}
Here $\rho _s$ is the normal state resistivity of a superconductor and $%
F=Y^{*}/Y$, where $Y^{*}$ represents the charge current in FS contact

\begin{equation}
Y^{*}=\sum_{\uparrow \downarrow }P_{\uparrow ,\downarrow }\int \frac{%
d^2k_{\parallel }}{(2\pi )^2}\int_{-\infty }^{+\infty }\left( -\frac{%
\partial f}{\partial E}\right) \left[ 1-C_{\uparrow ,\downarrow
}(E)-D_{\uparrow ,\downarrow }(E)\right] N_s^{-1}(E)dE  \label{Yc}
\end{equation}
and $Y$ represents the total current 
\begin{equation}
Y=\sum_{\uparrow \downarrow }P_{\uparrow ,\downarrow }\int \frac{%
d^2k_{\parallel }}{(2\pi )^2}\int_{-\infty }^{+\infty }\left( -\frac{%
\partial f}{\partial E}\right) \left[ 1+\frac{k_{F\downarrow ,\uparrow }}{%
k_{F\uparrow ,\downarrow }}A_{\uparrow ,\downarrow }(E)-B_{\uparrow
,\downarrow }(E)\right] dE.  \label{Y}
\end{equation}
Here $N_s^{}(E)=E/\sqrt{E^2-\Delta ^2(T)}$ is the density of states in a
superconductor.

At $E<\Delta (T)$ the coefficients $C,D$ vanish, while $A,B$ are given by

\begin{equation}
\frac{k_{F\downarrow ,\uparrow }}{k_{F\uparrow ,\downarrow }}A_{\uparrow
,\downarrow }(E)=1-B_{\uparrow ,\downarrow }(E)=\frac{T_{\uparrow
}T_{\downarrow }(1+\alpha ^2)}{(1-r_{\uparrow }r_{\downarrow })^2+\alpha
^2(1+r_{\uparrow }r_{\downarrow })^2},  \label{a,b}
\end{equation}
where $\alpha =\sqrt{\Delta ^2(T)-E^2}/E$. At $E>\Delta (T)$%
\begin{equation}
\frac{k_{F\downarrow ,\uparrow }}{k_{F\uparrow ,\downarrow }}A_{\uparrow
,\downarrow }(E)=\frac{T_{\uparrow }T_{\downarrow }(1-\beta ^2)}{\left[
1+\beta -r_{\uparrow }r_{\downarrow }(1-\beta )\right] ^2};\text{ }%
1-B_{\uparrow ,\downarrow }(E)=\frac{T_{\uparrow }T_{\downarrow }(1-\beta
)^2+4\beta (T_{\uparrow }+T_{\downarrow })}{\left[ 1+\beta -r_{\uparrow
}r_{\downarrow }(1-\beta )\right] ^2};  \label{AB}
\end{equation}

\begin{equation}
C_{\uparrow ,\downarrow }(E)=\frac{2(1+\beta )(T_{\uparrow }+T_{\downarrow })%
}{\left[ 1+\beta -r_{\uparrow }r_{\downarrow }(1-\beta )\right] ^2};\text{ }%
\frac{k_{F\downarrow ,\uparrow }}{k_{F\uparrow ,\downarrow }}A_{\uparrow
,\downarrow }(E)+B_{\uparrow ,\downarrow }(E)+C_{\uparrow ,\downarrow
}(E)+D_{\uparrow ,\downarrow }(E)=1,  \label{CD}
\end{equation}
where $\beta =\sqrt{E^2-\Delta ^2(T)}/E$.

The results of the calculations of the excess resistance factor $F=Y^{*}/Y$
for a FS contact are shown in Fig.5. It is seen that $F$ increases strongly
at temperatures close to $T_c$. Given the fact that the charge-imbalance
relaxation length $\lambda _Q$ becomes macroscopically large near $T_c$ \cite
{BTK,Hsiang,Ryaz} we conclude that measuring the excess resistance in a FS
contact can provide a sensitive probe for measuring spin polarization.

In conclusion, we have presented the results of a theoretical study of
interface resistance in ferromagnet/superconductor junctions. The Andreev
reflection theory is extended in order to take into account the impurity
scattering within the contact in the regime of strong disorder. The model is
applied to the calculation of the excess resistance of a FS contact caused
by penetration of an electric field into a superconductor. The latter
contribution could be important in contacts with planar geometry and
provides an additional method for measuring spin polarization in
ferromagnets.

{\bf Acknowledgments.} This work was supported in part by NWO program for
Dutch-Russian research cooperation. Stimulating discussions with J. Aarts,
G. Gerritsma, M.Yu. Kupriyanov, I.I. Mazin, B. Nadgorny and V.V. Ryazanov
are gratefully acknowledged.

\newpage

\begin{figure}
\par
\begin{center}
\mbox{\epsfxsize=0.8\hsize \epsffile{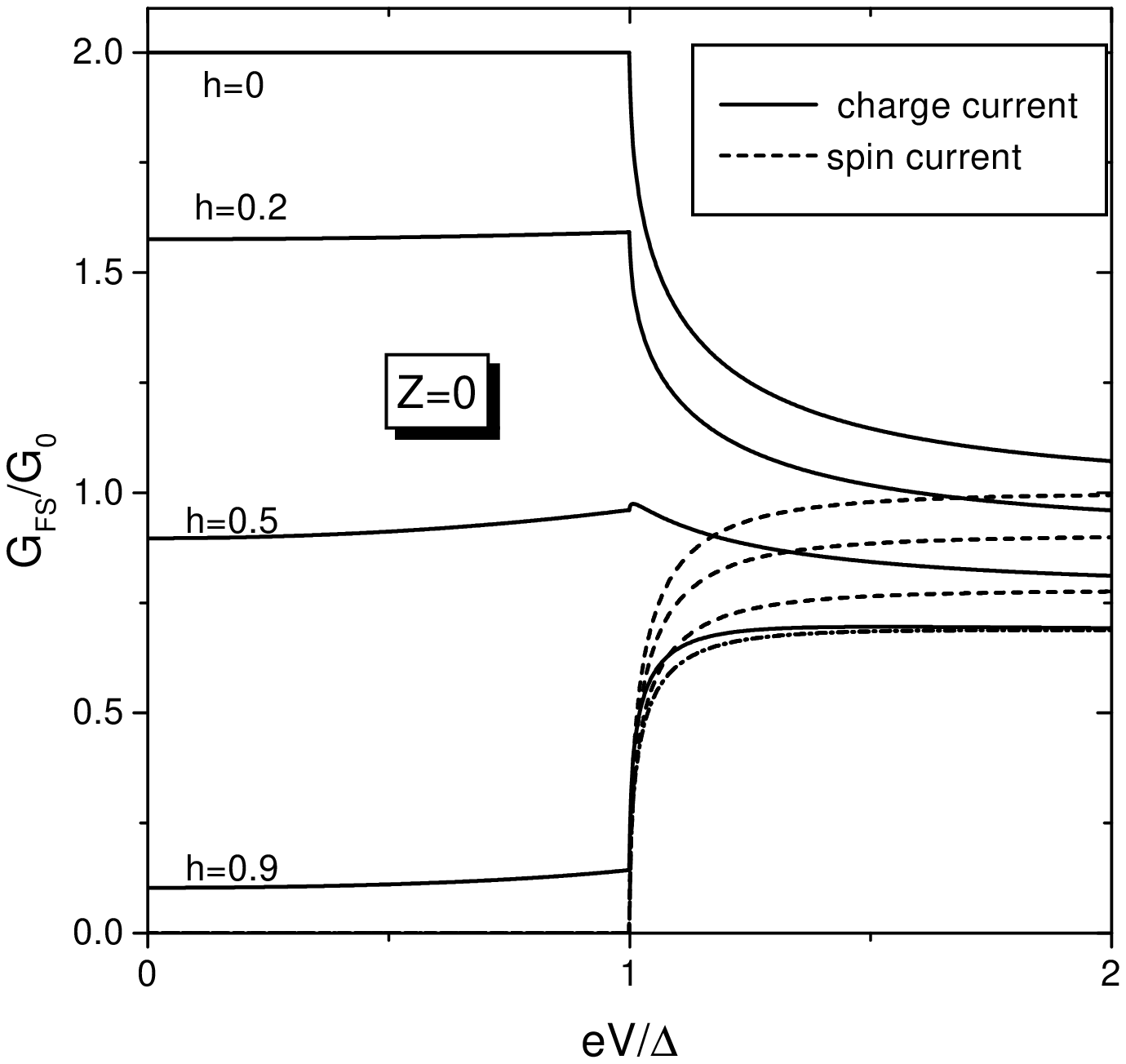}}
\end{center}
\caption{Low temperature spin and charge currents in a ballistic FS contact
for different spin polarizations in a ferromagnet for the barrier strength
parameter Z=0.}
\end{figure}

\begin{figure}
\par
\begin{center}
\mbox{\epsfxsize=0.8\hsize \epsffile{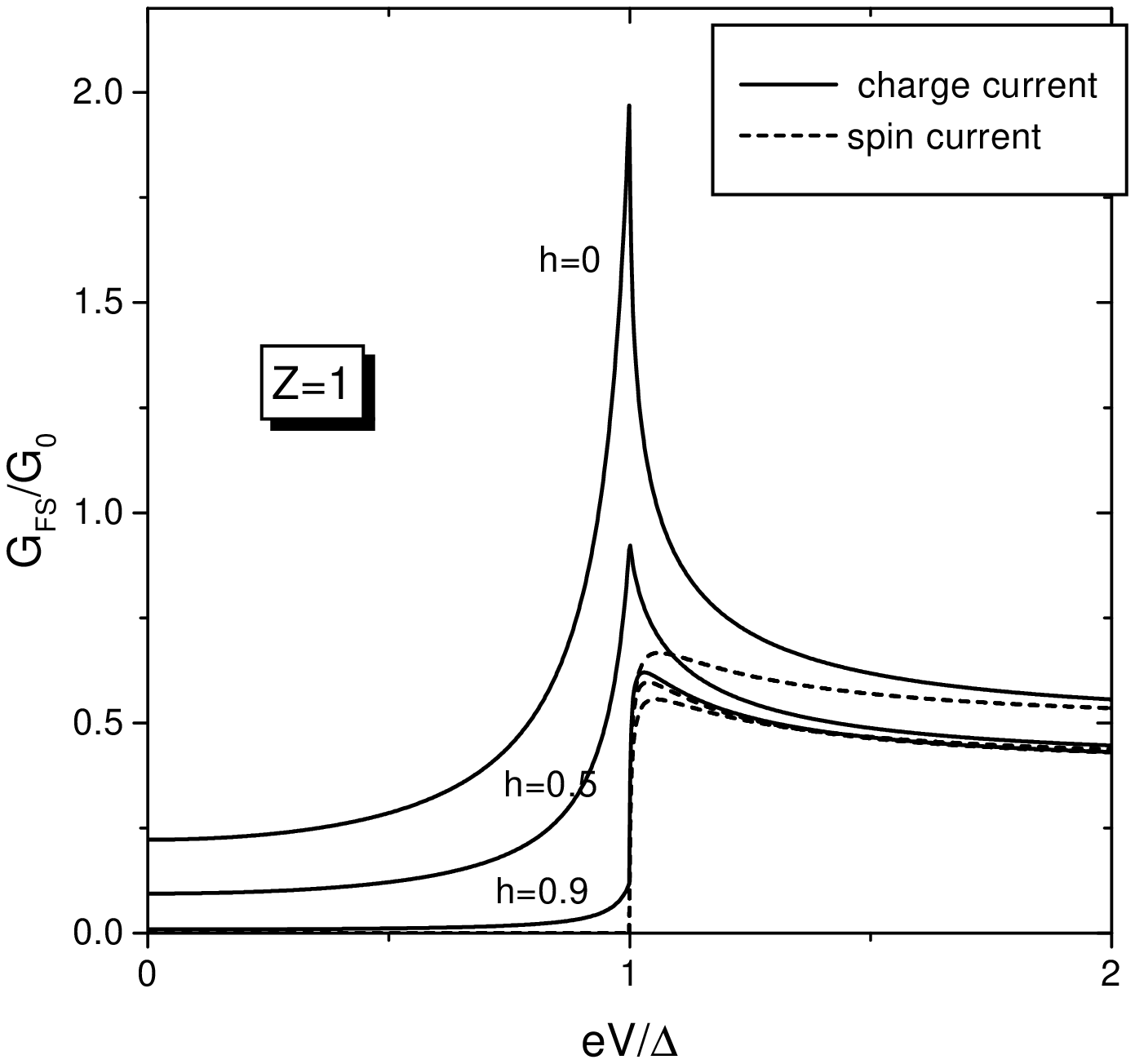}}
\end{center}
\caption{Low temperature spin and charge currents in a ballistic FS contact
for different spin polarizations in a ferromagnet for the barrier strength
parameter Z=1.}
\end{figure}

\begin{figure}
\par
\begin{center}
\mbox{\epsfxsize=0.8\hsize \epsffile{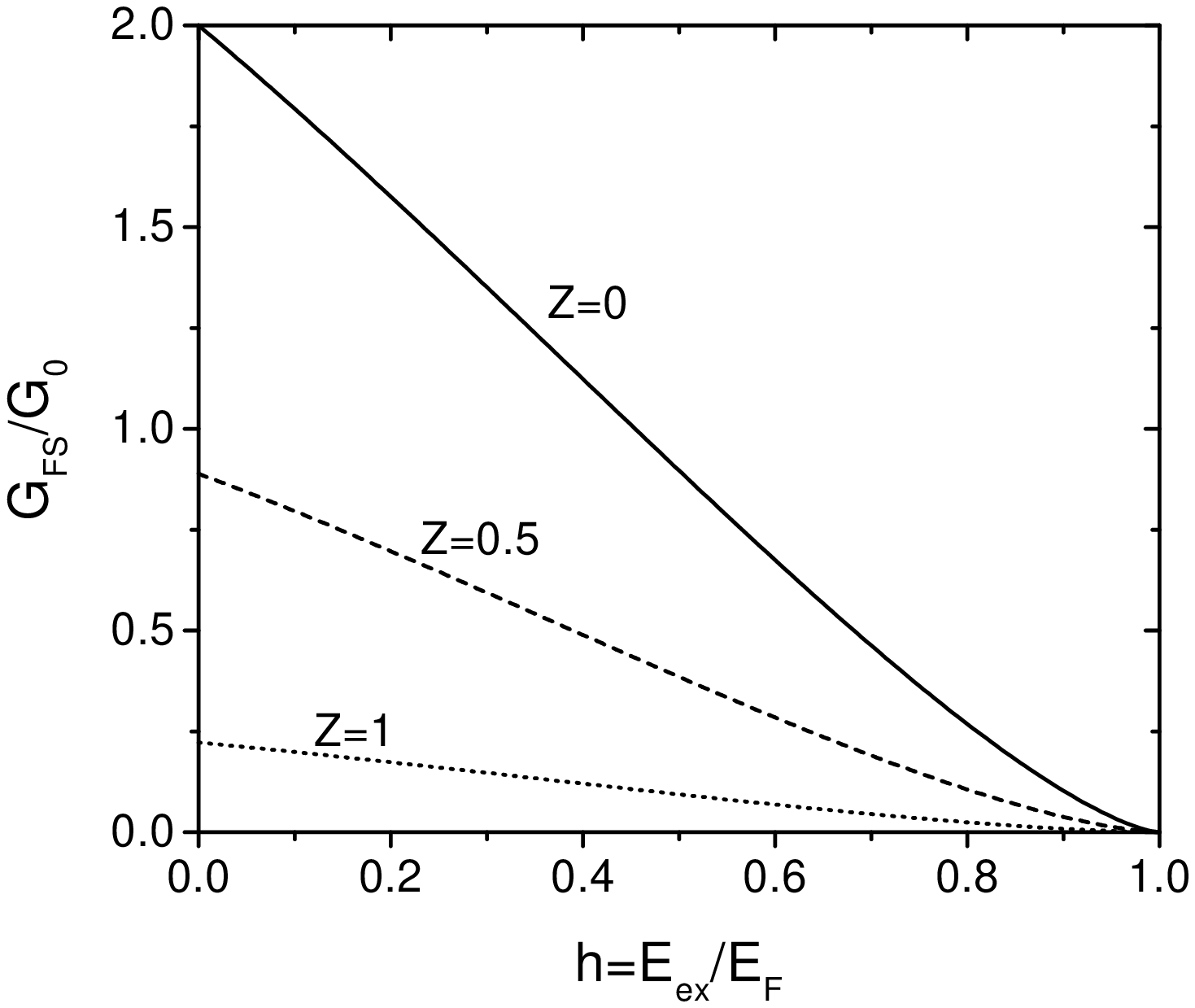}}
\end{center}
\caption{Zero-bias conductance of a ballistic FS contact as a function of the
spin polarization at various barrier strengths.}
\end{figure}

\begin{figure}
\par
\begin{center}
\mbox{\epsfxsize=0.8\hsize \epsffile{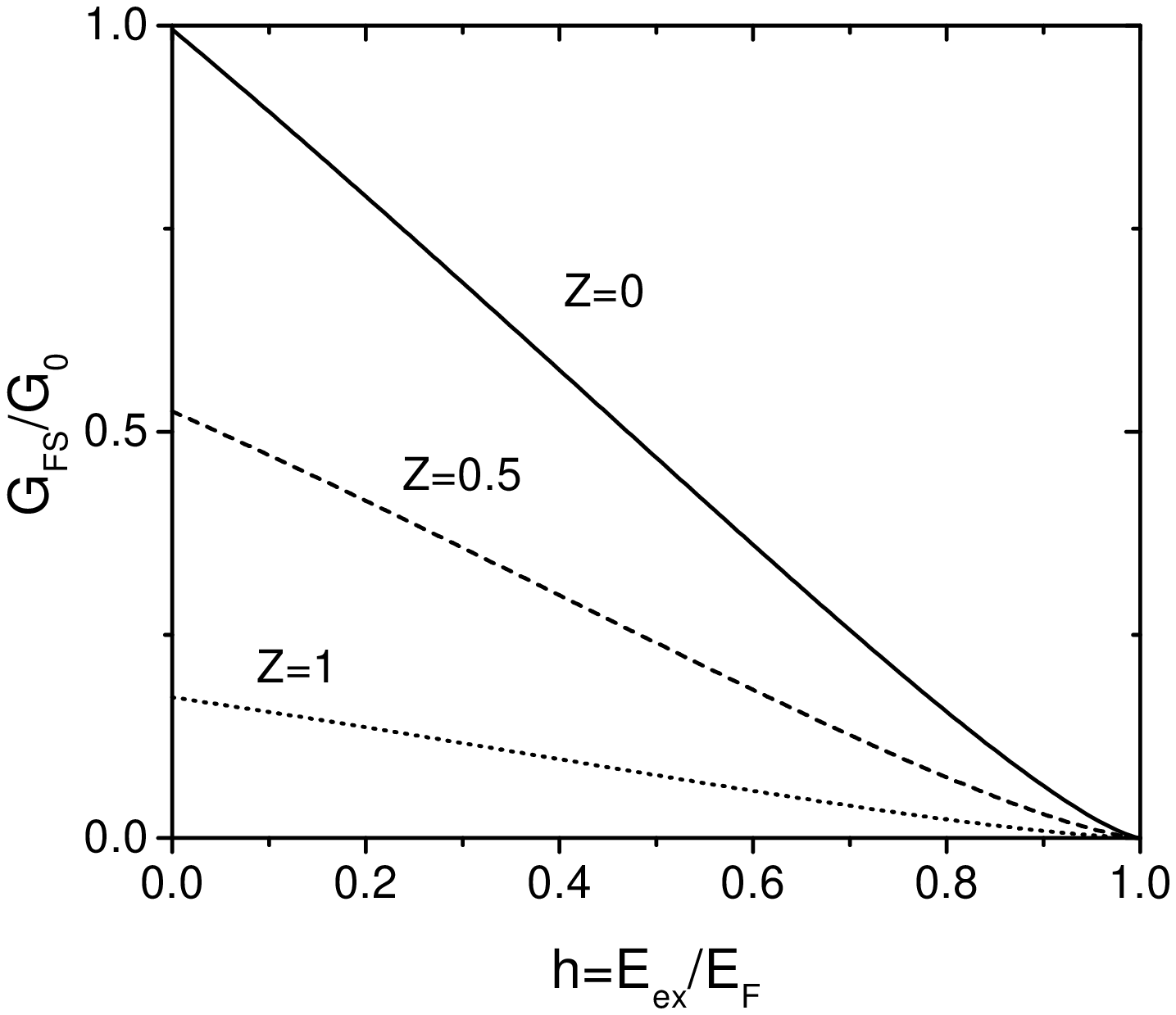}}
\end{center}
\caption{Zero-bias conductance of a disordered FS contact as a function of the
spin polarization at various barrier strength.
}
\end{figure}

\begin{figure}
\par
\begin{center}
\mbox{\epsfxsize=0.8\hsize \epsffile{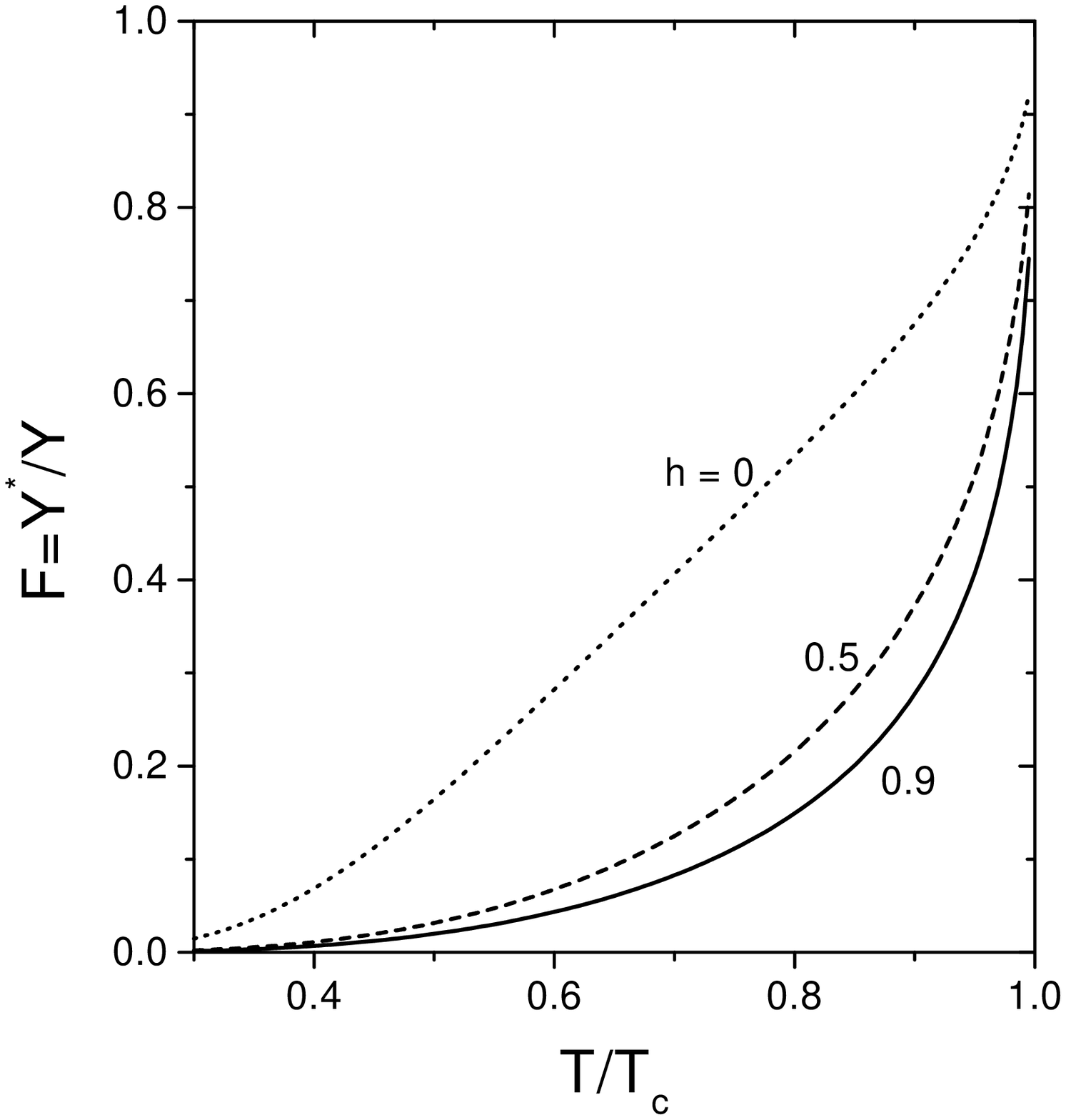}}
\end{center}
\caption{Temperature dependence of the excess resistance in a FS contact.}
\end{figure}

\end{document}